\documentclass{PHYEAUTH}

\usepackage{graphicx}
\usepackage{amsmath}
\usepackage{amssymb}

\begin{document}

\begin{frontmatter}

\title{Quantum interference in bilayer graphene}

\author[address1]{R.~V.~Gorbachev\thanksref{thank1}},
\author[address1]{F.~V.~Tikhonenko},
\author[address1]{A.~S.~Mayorov},
\author[address1]{D.~W.~Horsell},
and
\author[address1]{A.~K.~Savchenko}

\address[address1]{School of Physics, University of Exeter, Stocker Road, Exeter, EX4 4QL, UK}

\thanks[thank1]{Corresponding author. Fax: +44 1392264111, E-mail: Roman.Gorbachev@exeter.ac.uk}

\begin{abstract}
We report the first experimental study of the quantum interference correction to the conductivity of bilayer graphene. Low-field, positive magnetoconductivity due to the weak localisation effect is investigated at different carrier densities, including those around the electroneutrality region. Unlike conventional 2D systems, weak localisation in bilayer graphene is affected by elastic scattering processes such as intervalley scattering. Analysis of the dephasing determined from the magnetoconductivity is complemented by a study of the field- and density-dependent fluctuations of the conductance. Good agreement in the value of the coherence length is found between these two studies.
\end{abstract}

\begin{keyword}
weak localisation \sep magnetoconductance \sep bilayer systems \sep graphene
\PACS 73.23.-b \sep 72.15.Rn \sep 73.43.Qt
\end{keyword}

\end{frontmatter}

\section{Introduction}

The recently developed technique to fabricate free-standing sheets of graphene \cite{NovoselovS} brought with it the first opportunity to study a true two-dimensional crystal. Since this development, remarkable phenomena have been predicted and observed in both monolayer and bilayer forms of this crystal \cite{GeimNatMat}. One of the important questions that has arisen is the influence of the chiral nature of charge carriers in graphene on the quantum interference. In recent theories \cite{McCannPRL06,Kechedzhi}, this influence was considered on the quantum correction to the conductivity due to weak localisation (WL). This phenomenon is well studied in conventional 2D systems \cite{AltshulerPRB80}, but in graphene the quantum interference was predicted to be affected not only by inelastic scattering, but also by elastic inter- and intra-valley scattering. It was shown that the shape of the low-field magnetoconductivity (MC) of bilayer graphene \cite{Kechedzhi} can reflect the interplay between these scattering mechanisms.

Here we report an experimental investigation of the MC of bilayer graphene. We show how the different scattering processes are revealed in the weak localisation and how they depend on the carrier density.

\section{Experiment}

\begin{figure}[h]
\begin{center}\leavevmode
\includegraphics[width=.95\columnwidth]{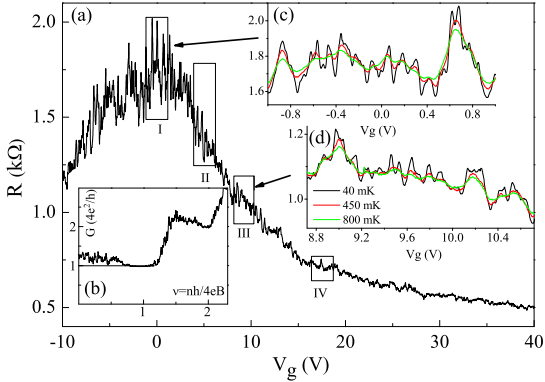}
\caption{(a) Field-effect in bilayer graphene. The numbered boxes indicate the regions where the magnetoconductivity was studied. (b) Quantum Hall effect in bilayer graphene. (c), (d) Temperature smearing of conductance fluctuations as a function of gate voltage (carrier density).}\label{fig:one}
\end{center}
\end{figure}

The sample was fabricated by the procedure developed in \cite{NovoselovS}. Highly-oriented pyrolytic graphite was mechanically exfoliated and deposited on top of a thermally oxidized $n^+$Si substrate (with an oxide thickness of $300$\,nm). A pair of Ohmic (Au/Cr) contacts was deposited using electron-beam lithography on opposite edges of a bilayer flake of length 1.5\,$\mu$m and width 1.8\,$\mu$m. The highly-doped substrate was used as a back gate and the carrier density in the flake was controlled by the gate voltage $V_g$. (Using the formula for a plane-plate capacitor, one can find a simple relation between the gate voltage and the carrier density: $dn/dV_g=7.3\times10^{10}$\,cm$^{-2}$V$^{-1}$.) Electrical measurements were performed in the temperature range from 0.04\,K to 4.2\,K. The gate voltage dependence of the resistance of the studied sample is presented in Fig.~\ref{fig:one}(a). Far from electroneutrality region at $V_g=0$, where the charge carriers change from electrons to holes, the sample exhibits an almost constant mobility of $\approx 7000\,$cm$^2/$V$\cdot$s.

To verify that we are dealing with a true bilayer graphene system the quantum Hall effect was studied. The results are shown in Fig.~\ref{fig:one}(b). The values of the conductance at the plateaux correspond to those of a conventional 2D system with four-fold degeneracy; however, the filling factor $\nu$ at which the first plateau occurs is unconventional, and is a result of the presence of the lowest Landau level at zero energy. This is a clear demonstration that the sample is bilayer graphene \cite{GeimNatMat}.

\begin{figure}[h]
\begin{center}\leavevmode
\includegraphics[width=.95\columnwidth]{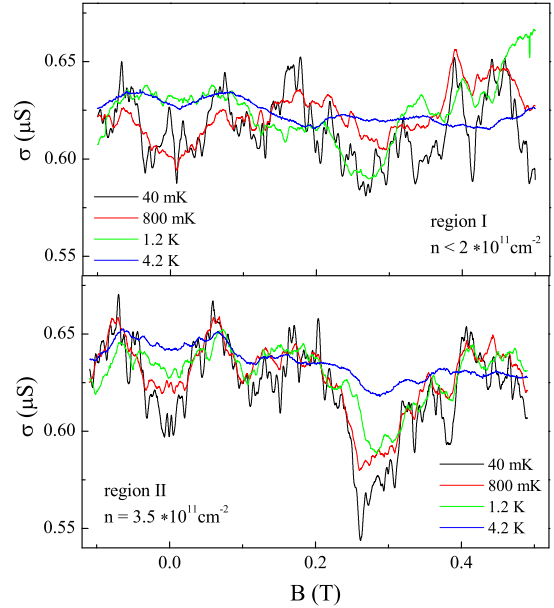}
\caption{Conductivity of the sample at low magnetic field near electroneutrality region (region I, top panel) and at higher concentration (region II, bottom panel).}\label{fig:two}
\end{center}
\end{figure}

\begin{figure}[h]
\begin{center}\leavevmode
\includegraphics[width=.9\columnwidth]{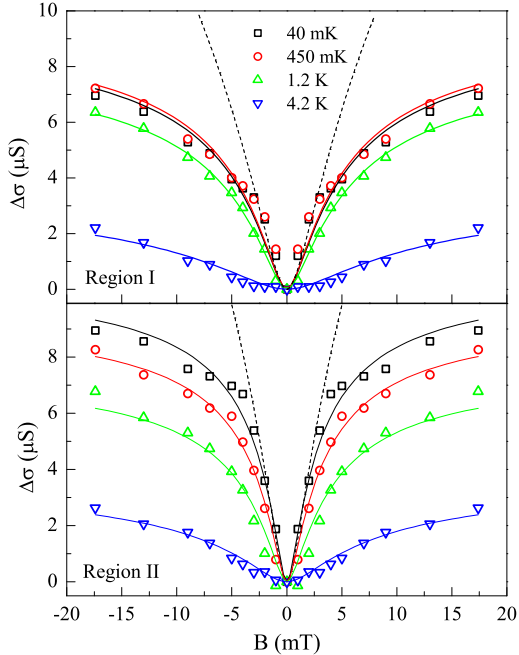}
\caption{Low-field magnetoconductivity of bilayer graphene. Solid lines are fits to the data by Eq.~\ref{eqn:one}. The dashed line is a fit by the same equation if only the first term is considered.}\label{fig:three}
\end{center}
\end{figure}

The low-field MC of the sample is strongly affected by fluctuations of the conductance, seen in Fig.~\ref{fig:one}. Since the size of the sample is rather small, these reproducible, mesoscopic fluctuations are significant. We observe fluctuations with amplitude $\sim e^2/h$, both as a function of the gate voltage (Fig.~\ref{fig:one}(c,d)) and magnetic field (Fig.~\ref{fig:two}). This amplitude decreases with increasing temperature, Fig.~\ref{fig:one}(c,d), which is comparable to the decrease of the amplitude of the MC. Therefore, straightforward analysis of the raw MC data is not possible and we have to apply an averaging procedure. We averaged the effect of the magnetic field on the conductivity across each of the studied regions, highlighted by boxes in Fig.~\ref{fig:one}(a) ($\Delta V_g = 2$\,V). (A detailed description of this averaging procedure is given in \cite{Gorbachev}.) Results of the averaging are presented in Fig.~\ref{fig:three} where the low-field MC in the electroneutrality region and electron region is shown. The experimental points are fitted by the theoretical expression \cite{Kechedzhi}:

\begin{eqnarray}\label{eqn:one}
\Delta\sigma(B)=\frac{e^2}{\pi h}\left[F\left(\frac{\tau_B^{-1}}
{\tau_{\phi}^{-1}}\right)-F\left(\frac{\tau_B^{-1}}{\tau_{\phi}^{-1}+
2\tau_{i}^{-1}}\right) \right.\nonumber\\
\left. +2F\left(\frac{\tau_B^{-1}}{\tau_{\phi}^{-1}+
\tau_{i}^{-1}+\tau_{*}^{-1}}\right)\right]\; .
\end{eqnarray}
Here $F(z)=\ln{z}+\psi{\left(0.5 + z^{-1} \right)}$, $\psi(x)$ is the digamma function, $\tau_B^{-1}=4eDB/\hbar$, $\tau_\phi^{-1}$ is the phase-breaking rate and $\tau_*^{-1}=\tau_z^{-1}+\tau_w^{-1}$, where $\tau_w$ is the intra-valley ``warping'' time and $\tau_z$ is the time of chirality breaking \cite{MorpurgoPRL06}. It is seen from Eq.~\ref{eqn:one} that the magnetoconductivity is affected not only by inelastic scattering ($\tau_\phi^{-1}$) but also by elastic scattering rates $\tau_i^{-1}$ and $\tau_*^{-1}$.

The solid lines in Fig.~\ref{fig:three} present the best fit of the experimental points by the first and second terms of Eq.~\ref{eqn:one}. The third term is found to be significantly smaller than the other two and produces only a minor correction to the fit. The second term plays a crucial role: the saturation of the WL correction at higher fields is due solely to this term. If only the first term is considered, the MC increases continuously with increasing magnetic field. This is shown by the dashed line in Fig.~\ref{fig:three}, which is a best fit to the low-field data by the first term alone.

\section {Discussion}

\begin{figure}[h]
\begin{center}\leavevmode
\includegraphics[width=.95\columnwidth]{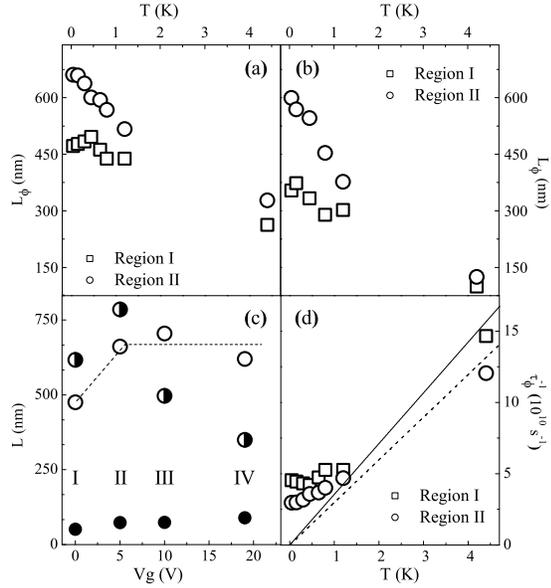}
\caption{(a) Coherence length found from the WL analysis for regions I and II. (b) Values of $L_\phi$ found from the analysis of the amplitude of the conductance fluctuations as a function of the magnetic field. (c) Comparison of the characteristic lengths in bilayer graphene at different concentrations: saturation value of $L_\phi$ (open circles), intervalley scattering length $L_i$ (half-filled circles) and mean free path $l_p$ (filled circles). (The dashed line is is a guide to the eye to show the decrease of the saturation value of $L_\phi$ in the electroneutrality region.) (d) Temperature dependence of the dephasing rates in the same regions. Lines are drawn according to Eq.~\ref{eqn:two}.}\label{fig:four}
\end{center}
\end{figure}

Figure~\ref{fig:four} contains the main results of our analysis. Figure~\ref{fig:four}(a) presents the temperature dependence of the coherence length $L_\phi$ for the electroneutrality and electron regions. At low temperatures, $L_\phi$  saturates at a value comparable to the size of the sample. This saturated value is smaller in the electroneutrality region than in the electron region. A possible reason for this is that in the electroneutrality region the carrier density, which is on average zero, becomes inhomogeneous due to the formation of electron--hole puddles. This results in an effective reduction of the size of the sample.

We analyse the fluctuations of the conductance in terms of the conventional theory of universal conductance fluctuations \cite{LeePRL55}. From the amplitude of these fluctuations, averaged over the magnetic field range $\Delta B=0.5\,\mathrm{T}$, we determine the coherence length as a function of the temperature, Fig~\ref{fig:four}(b). These values of $L_\phi$ agree well with the results of the WL analysis. The elastic scattering in bilayer graphene therefore appears to affect the fluctuations of the conductance less than the magnetoconductivity due to WL.

Figure~\ref{fig:four}(c) shows the relation between the characteristic lengths in bilayer graphene: the saturation value of $L_\phi$, the intervalley scattering length $L_i$, and the mean free path $l_p$. As mentioned above, the saturation value of $L_\phi$ is seen to be somewhat smaller in the electroneutrality region than at higher densities, as has also been seen in monolayer graphene systems \cite{Tikhonenko}. In all studied regions, $L_i$ is found to be temperature independent and comparable to the saturation value of $L_\phi$. This implies that, similar to \cite{Tikhonenko}, there is a significant contribution of the sample edges to intervalley scattering. There is a trend of a decrease of $L_i$ with increasing density; but its origin is not clear, as, contrary to single-layer, the density of states in bilayer graphene is energy independent. Both $L_\phi$ and $L_i$ are found to be significantly larger than the mean free path $l_p$, which justifies the application of the diffusive theory \cite{Kechedzhi}. The small contribution of the third term in Eq.~\ref{eqn:one} is in agreement with theoretical expectations \cite{Kechedzhi}: $\tau_w \approx \tau_p$ for the studied range of densities.

The temperature dependence of the dephasing rate is presented in Fig.~\ref{fig:four}(d). At temperatures $\geq 1$\,K it can be described by the conventional theory \cite{AltshulerPRB80} for dephasing due to the electron--electron interaction in the diffusive regime, $T\tau_p\ll 1$:
\begin{equation}\label{eqn:two}
\tau_{\phi}^{-1} = \frac{\beta k_BT}{\hbar g}\ln{g}\; ,
\end{equation}
where $g=\sigma/(e^2/h)$ is a dimensionless conductivity, and $\beta$ is an empirical coefficient close to unity.

\section {Conclusion}

The quantum correction to the conductivity due to weak localisation in bilayer graphene is affected both by inelastic and elastic scattering mechanisms. This is caused by the chirality of charge carriers in this system. Values of the coherence length and intervalley scattering length are determined from analysis of the WL. The found values of the coherence length are in agreement with those determined from analysis of the observed conductance fluctuations as a function of magnetic field. The latter is done in terms of the conventional, unmodified theory of universal conductance fluctuations. The temperature dependence of the dephasing rate at high temperatures can be described by the theory of electron--electron interaction in the diffusive regime.

\end{document}